# Adaptive dynamic range shift (ADRIFT) quantitative phase imaging

Keiichiro Toda,[1] Miu Tamamitsu,[1] and Takuro Ideguchi[1,2,3,*]

[1]Department of Physics, The University of Tokyo, Tokyo 113-0033, Japan
[2]Institute for Photon Science and Technology, The University of Tokyo, Tokyo 113-0033, Japan
[3]PRESTO, Japan Science and Technology Agency, Saitama 332-0012, Japan
[*]Corresponding author: ideguchi@ipst.s.u-tokyo.ac.jp

**Quantitative phase imaging (QPI) is often used for label-free single cell analysis with its high-contrast images of optical phase delay (OPD) map. Contrary to other imaging methods, sensitivity improvement has not been intensively explored because conventional QPI is sensitive enough to see surface roughness of a substrate which anyway restricts the measurable minimum OPD. However, emerging QPI techniques which utilize, for example, differential image analysis of consecutive temporal frames, such as mid-infrared photothermal QPI, mitigate the minimum OPD limit by decoupling the static OPD contribution and allow to measure much smaller OPD. Here, we propose and demonstrate super-sensitive QPI with expanded dynamic range. It is enabled by adaptive dynamic range shift with combination of wavefront shaping and dark-field QPI techniques. As a proof-of-concept demonstration, we show dynamic range expansion (sensitivity improvement) of QPI by a factor of 6.6 and its utility for improving sensitivity of mid-infrared photothermal QPI. This technique can also be applied for wide-field scattering imaging of dynamically changing nanoscale objects inside and outside a biological cell without losing global cellular morphological image information.**

Phase imaging[1-17] provides morphological phase contrast of transparent samples and is widely used in various fields especially in biological science because morphological features of micrometer-scale specimens give us valuable information of complex biological phenomena. Quantitative phase imaging (QPI)[1-6] is the most powerful method to study cellular morphology among various phase imaging methods such as phase-contrast[7] and differential-interference-contrast[8] imaging because it is able to accurately measure the optical phase delay (OPD) caused by a sample. This quantitative nature enables, for example, cellular dry-mass and growth-rate analyses[9], which have been recognized as new tools for single cell analysis. Since the detectable minimum OPD of conventional QPI (~10 mrad) is already small enough to clearly observe surface roughness of a glass slide that limits the detectable sample-induced OPD, sensitivity improvement of QPI has not been intensively explored. However, recent development of pump-probe-type perturbative QPI techniques such as mid-infrared (MIR) photothermal QPI[10-13] have shown that temporal differential analysis of consecutively measured images can cancel out the substrate's background and reveal small OPD change. Thus, sensitivity improvement of QPI is highly demanded in this context. In addition, wide-field scattering imaging techniques using dark-field (DF)[14,15] or interferometric scattering (iSCAT)[16,17] have also used the concept of differential image analysis to observe dynamically changing small signals from nanoscale scattering objects. They have been mostly used for investigating simple bio-mimicking systems[16,17], and more recently, applied to measure gold nanoparticles on cell membranes[18,19]. However, the limited dynamic range of DF causes decreased sensitivity under the existence of highly scattering medium such as cells. On the other hand, iSCAT is not a technique to quantitatively and comprehensively measure a complex structure of specimens because it is only sensitive to the medium boundaries. These features prevent simultaneous detection of quantitative global cellular structure and the small scattering signals. To measure dynamic motion of small particles such as exosomes, liposomes and viruses inside and outside a living cell, we need to detect small signals on top of large background from the cell with higher measurement dynamic range in a manner of QPI.

Conventional QPI techniques, however, are not made to detect small OPD because strong unscattered light (also known as zeroth order diffracted light), which brings no information of sample's morphology, dominantly exposes the image sensor and its shot noise determines dynamic range of the measurement. With a commonly used image sensor, the measurable OPD range is from ~10 mrad to ~1 rad in a shot-noise-dominant imaging condition, limiting the sensitivity of the system[6] (Fig. 1a). The sensitivity can be improved by increasing the photon flux of the scattered light reaching the image sensor, because it contains the morphological information of the sample. This is realized by dark-field (DF) imaging[14,15], where the undesired strong unscattered light is rejected with a spatial filter in the Fourier plane. Its dynamic range can be shifted to the smaller-OPD regime by increasing the illumination light up to the level where the brightest spot caused by the largest OPD in the field of view (FOV) nearly saturates the image sensor. However, the sensitivity improvement cannot be significant when the sample has a global structure that causes large OPD (~1 rad) because the dynamic range is "pinned" to this value (Fig. 1a). Therefore, the DF imaging technique is also not applicable to observe small OPD changes embedded in microscale large OPD structure. For example, single cell imaging falls into this situation in any cases. To date, any existing phase imaging techniques, not only DF imaging, fail to simultaneously measure subtle OPD changes and large OPD because dynamic range is pinned to the larger side. To address the challenge, it is necessary to expand dynamic range of the OPD measurement.

Here, we propose and demonstrate a method to expand dynamic range of phase imaging, which we call adaptive dynamic range shift quantitative phase imaging (ADRIFT-QPI). It works by separately measuring the sample's large and small OPD distributions and seamlessly connecting them (Fig. 1a). In addition to measuring large OPD with the conventional QPI, we measure small OPD with a new technique called phase-canceling dark-field quantitative phase imaging (PC-DF-QPI), which is enabled by using a wavefront shaping technique in a configuration of dark-field QPI (DF-QPI). Our proof-of-concept experiments demonstrate dynamic range expansion (sensitivity improvement) of QPI measurement by a factor of 6.6, which corresponds to 44 times speed improvement, and show significant sensitivity improvement of MIR photothermal QPI. The concept of this technique will promise the remarkable advancement of label-free imaging with its high sensitivity.

**Results**

**Principle of ADRIFT-QPI.** The working principle of the dynamic range expansion is illustrated in Fig. 1b. In the first measurement, a large OPD distribution is measured by conventional QPI with a plane wave illumination. Then the large OPD distribution is optically canceled by wavefront shaping with a phase-only spatial light modulator (SLM) [20-22] such that the light returns to be a quasi-plane wave which can be focused around a single spot in the Fourier plane where a single dot spatial mask (DF mask) selectively rejects the focused light. The DF spatial mask allows only a small amount of light, which deviates from the plane wave, to reach the image sensor so that illumination photon flux on the sample can be increased without saturating the sensor. Thus, in the second measurement performed by PC-DF-QPI with stronger illumination light, dynamic range of the measurement is shifted to the smaller-OPD regime. The sensitivity of PC-DF-QPI determines that of ADRIFT-QPI. Note that, to maintain phase quantitativity, we implement a technique to perform QPI in the DF configuration (DF-QPI). Finally, a dynamic-range-expanded OPD image of the sample is computationally reconstructed by seamlessly connecting the SLM's input OPD map for phase canceling and the measured PC-DF-QPI image. To guarantee high sensitivity, it is needed to precisely calibrate the response of the SLM with respect to the input numerical value, so that the SLM's phase ambiguity becomes lower than that offered by the sensitivity of the PC-DF-QPI. Note that the OPD image obtained in the first measurement cannot be used for the computational reconstruction due to the digitization noise of the SLM (i.e., B≠B' in Fig. 1b).

**Optical layout of ADRIFT-DH.** The optical implementation of the system is presented in Fig. 1c. In this work, we implement off-axis digital holography (DH) as a QPI technique (therefore, we replace -QPI by -DH, below). Other QPI techniques can also be applied to this concept in general[2,4]. DH and DF-DH is interchangeable in a single setup by insertion/removal of a DF mask placed in the Fourier plane. A phase-only SLM is put in a sample's conjugate plane for phase canceling[23]. To measure a sample whose large-OPD structure does not move during the imaging frames, the SLM's input OPD is not required to be refreshed for every measurement, meaning no need to switch the system between DH and DF-DH. Off-axis reference light is illuminated on the image sensor so that complex-field can be measured. The reference light also works as a local oscillator for heterodyne detection to amplify the signal and guarantee shot-noise-limited measurements, which is especially important in PC-DF-DH measurements where object light is significantly reduced. Note that, to the best of our knowledge, this work is the first demonstration of QPI in DF configuration (DF-QPI). The specification of the optical system is given in Methods.

**OPD sensitivity of ADRIFT-DH.** We theoretically discuss the OPD sensitivity of PC-DF-DH, which determines that of ADRIFT-DH. For simplicity, we assume a case where a transparent sample is illuminated by a plane wave with uniform amplitude distribution $U_0$. The complex amplitude of the light from the sample arm in DH configuration at the image sensor can be written as $U_0 e^{i\theta_{mn}}$, where $\theta_{mn}$ denotes the OPD map introduced by the sample ($m$ and $n$ are indices of the image sensor's pixels along $x$ and $y$ directions, respectively). Intensity at the image sensor in DF imaging with a DC-cut spatial mask placed in the Fourier plane can be approximated to be $\left|U_0(e^{i\theta_{mn}^{PC}} - 1)\right|^2$ when the amount of unscattered light does not largely change with and without the sample. If the maximum OPD in the field of view (FOV) after phase canceling is sufficiently small ($\theta_{max}^{PC} \ll 1$), the maximum DF intensity found in the FOV may be described as

$$\left|U_0(e^{i\theta_{mn}^{PC}} - 1)\right|^2 = 2U_0^2(1 - \cos\theta_{max}^{PC}) \sim U_0^2 {\theta_{max}^{PC}}^2. \tag{1}$$

Therefore, in PC-DF-DH, we can increase the amount of illumination light to the sample by a factor of $1/{\theta_{max}^{PC}}^2$ because DH intensity provided by the sample arm is $\left|U_0 e^{i\theta_{mn}}\right|^2 = U_0^2$ for any samples. The stronger illumination light leads to dynamic range shift to the smaller-OPD regime with $\sim 1/\theta_{max}^{PC}$ times reduction of the detected shot noise of illumination light (sensitivity improvement). In addition, there is another but minor factor of sensitivity improvement: shot-noise reduction by the phase canceling itself. By canceling the OPD distribution due to the sample, the amount of light from the sample arm can be reduced down to half before canceling, at most. This factor can reduce the shot noise by a factor between 1-1.4, which depends on the amount of canceled OPD, in addition to the above mentioned improvement factor of $1/\theta_{max}^{PC}$ from the stronger illumination.

**Experimental validation of DF-DH.** We provide experimental validation of DF-DH which is used as a QPI technique in ADRIFT-DH. Figures 2a and 2b confirm QPI capability of DF-DH, where the OPD distribution of a 5 μm silica microbead measured by DF-DH shows good agreement with that obtained with conventional DH. The DF-DH image is reconstructed by using an image of the sample-specific scattered light measured by DF-DH and that of the unscattered light without the sample measured by DH.

We next confirm that the smaller maximum OPD value in the FOV allows us to increase the illumination intensity to the sample. A virtual object with arbitrary OPD values (0.16, 0.33 or 0.56 rad) is created with the SLM (indicated by the arrow in Fig. 2c) and measured with DF-DH by adjusting the illumination intensity to the sample to use the full dynamic range of the image sensor. The illumination intensity ratio of DF-DH to DH as a function of the maximum OPD is plotted in Fig. 2d. The measured data are in good agreement with the theoretical values, $1/\theta_{max}^2$, derived in Eq. (1).

Finally, we evaluate how the noise (temporal standard deviation of OPD) of DF-DH depends on the intensity of the light illuminating the sample. The noise is evaluated by taking standard deviation of 25 continuously measured temporal data points acquired at 10 Hz and averaging over 60 × 80 pixels (~30 μm × 40 μm) within the area indicated by the red rectangle in Fig. 2c. As Fig. 2e shows, the evaluated data are in good agreement with the theoretical curve and the noise is decreased down to 1.0 mrad with 33 times higher illumination intensity. As a reference, the noise of DH measurement is also shown in Fig. 2e. It stays at 6.0 mrad for any samples because illumination light is kept at the same amount. This is because the OPD produced by a transparent object appears as the spatial shift of the interference fringes rather than the change in the optical intensity.

**Experimental validation of phase canceling.** As discussed above, the amount of dynamic range shift to the smaller-OPD regime is determined by the maximum OPD value after phase canceling. We validate phase canceling method by measuring a large-OPD object (5 μm silica microbead) in Fig. 3. Figure 3a shows OPD images measured by DH before and after phase canceling. The OPD value of the microbead (~1 rad) is well canceled down to less than 0.1 rad, showing the phase canceling concept works. Figure 3b shows DF intensity images of the silica microbead provided by the sample-arm before and after phase canceling. Figure 3c shows an intensity image of the sample-arm in DH as a reference. In Figs. 3b and 3c, the maximum image sensor count is reduced from ~1,100 to ~30, a factor of ~35, by switching the system from DH to PC-DF-DH. The comparison clearly shows the significant reduction of the amount of light exposed on the image sensor by PC-DF-DH. It enables us to increase the illumination light to the sample and improve the OPD sensitivity.

**Dynamic-range-expanded MIR photothermal QPI of microbeads.** To illustrate the advantage of ADRIFT-QPI, we apply this technique to MIR photothermal QPI[10-13]. MIR photothermal QPI is a recently developed molecular vibrational imaging method, where the refractive-index change of the sample caused by the absorption of MIR pump light is detected through the OPD change of visible probe light. In this experiment, silica microbeads immersed in refractive-index matching oil are used as a sample. The MIR pump pulsed laser is tuned to the wavenumber of 1,045 cm$^{-1}$ which is resonant to the O-Si-O stretching mode of silica. Figure 4a shows the pump-OFF-state OPD images obtained by conventional DH and ADRIFT-DH. We can see the same OPD images including the background surface roughness of the glass plate. The slight deviation between the images comes from the highpass filtering effect of the DF mask used in ADRIFT-DH. Figure 4b shows the OPD change (pump ON-OFF) due to absorption of the MIR pump light, and Fig. 4c is the cross-section of the microbead images. In ADRIFT-DH measurement, the probe light illumination to the sample is 38 times higher than that of DH measurement by decreasing the maximum OPD value down to ~0.1 rad by phase canceling. It reduces the noise floor by a factor of 6.6. The small OPD change of a few mrad can be clearly visualized in ADRIFT-DH, which is otherwise buried in the optical shot noise in DH. This demonstration clearly shows advantage of the expanded dynamic range: capability of visualizing the original large-OPD (> 1 rad) distribution of the sample concurrently with the small OPD change (~mrad).

**Dynamic-range-expanded MIR photothermal QPI of a live biological cell.** As a more practical demonstration, we show dynamic-range-expanded MIR photothermal QPI of a live biological cell. The MIR pump light is tuned to 1,550 cm$^{-1}$ which is resonant to the peptide bond's amide II band mainly found in proteins. Figure 5a shows the pump-OFF-state OPD images obtained by DH and ADRIFT-DH. We decrease the maximum OPD value in the FOV to ~0.1 rad by phase canceling and increase the probe illumination light by a factor of 17, which is limited by the laser power in this particular case. The illumination light can be further increased with a proper light source to make full use of the sensor's dynamic range. The increased illumination light reduces the noise floor by a factor of 3.7. Figure 5b shows the OPD change between the ON and OFF states due to absorption of the MIR pump light. Only ADRIFT-DH clearly visualizes the signal localizations especially at the nucleoli and some particles indicated by the arrows in Fig. 5a, which could represent the richness of proteins[13].

Discussions

The amount of dynamic range shift (expansion) can be limited by two factors. One is the maximum OPD after phase canceling, and the other is the amount of increased light exposed on the sample. In our experiment, the OPD was canceled down to 0.1 rad, which allows us to increase the illumination light by a factor of 100, but in reality, it was limited to 38 because of the imperfection of the DF filtering due to wavefront distortion of the illumination light of the system, which can be mitigated with careful implementation of the system. The OPD can be further canceled by improving alignment of the SLM with respect to the sample's magnified image.

Theoretically, an 8-bit SLM allows for the OPD cancelation down to 0.025 rad such that ~1,000 times larger illumination, hence ~33 times higher sensitivity, is achievable. A larger-bit SLM would even improve it, although its noise would eventually be the limitation.

Our system can also be improved in several directions. The dynamic range can be better shifted with PC-DF-DH by implementing amplitude canceling in addition to phase canceling. It is especially useful for measuring non-transparent and/or defocused samples because amplitude distribution causes inefficient DF rejection. In addition, using a CMOS image sensor with ultra-high full well capacity[24] can further expand the dynamic range. The optical throughput can be improved by placing the SLM at the conjugate plane before the sample, which reduces the amount of illumination light, hence photodamage, to the sample[25].

In this work, we used ADRIFT-QPI for improving MIR photothermal QPI, but it can also be used for other applications. For example, there are some situations where the substrate's static roughness can be decoupled from the signal, such as flow cytometry[26], optical tweezer applications[27], optical diffraction tomography[28-30] and detection of dynamic OPD changes[6,9,12]. In addition, due to the capability to adaptively shift the dynamic range regardless of the sample condition, ADRIFT-QPI has a potential to be as sensitive as the state-of-the-art wide-field scattering imaging such as iSCAT, which is used for small particle measurement with extremely high sensitivity, even at the existence of highly scattering objects. We note that ADRIFT-QPI can be understood as a forward scattering counterpart of backscattering-based iSCAT. Therefore, it could bring a new opportunity to study behavior of small particles inside and outside cells without losing cellular morphological information. MIR photothermal QPI technique can also be implemented in the same system to add molecular contrast.

**Methods**

**Light source.** The visible light source is second harmonic generation (SHG) of a 10-ns, 1,000-Hz, 1,064-nm pulsed Q-switched laser (NL204-1K, Ekspla) with a nonlinear crystal LBO (Eksma Optics). The spatial mode of the SHG beam is cleaned by a single-mode optical fiber (P3-405B-FC-5, Thorlabs). The spectral bandwidth is ~2 nm after the fiber, which reduces the coherent noise. We note that a CW laser can be used for many applications of this technique, although a nanosecond pulsed laser is required as the probe light for MIR photothermal QPI.

**ADRIFT-DH system.** Linearly polarized light is created by a polarizer and its polarization direction is precisely adjusted by a half-wave plate to the orientation of liquid crystals in the SLM. The light is split into two by a beamsplitter (BS061, Thorlabs). In the sample arm, intensity of the illumination light can be adjusted with a neutral density (ND) filter (NDC-50C-2-A, Thorlabs) placed before the sample. The sample's image is magnified by a factor of 44 at the image sensor (acA2440-75μm, Basler) with an objective lens with NA 0.6 (LUCPLFLN40X, Olympus) and relay lenses (AC508-100-A and AC508-200-A, Thorlabs). The image sensor (acA2440-75μm, Basler) has a full well capacity of ~10,000 e$^-$/pixel. A phase-only SLM [1920×1152 XY Phase Series Spatial Light Modulator (Meadowlark Optics)] is placed in the sample's conjugate plane. A circular mask deposited on a glass substrate (50 or 100 μm in diameter, TOPRO) is put in the Fourier plane as a DF mask. In the reference arm, a delay line and a beam expander (A397TM-A and AC254-075-A, Thorlabs) adjust the differences in optical path length and beam diameter between the two arms, respectively. A transmission grating with 100 line-pairs/mm (66-341, Edmund Optics) placed in the sample's conjugate plane and two lenses (AC508-100-A, Thorlabs) create off-axis reference light. The laser's intensity fluctuation (1 - 2% in our case) is numerically compensated by recording a part of the light with another camera (acA2440-75μm, Basler) so that the shot-noise-limited detection is achieved. We mitigate background OPD fluctuation caused by the convection of the ambient air by enclosing the system with a box. The image sensor is operated at 10 or 20 Hz with the exposure time of 60 or 30 ms for the experiments shown in Figs. 2 – 4 or Fig. 5, respectively. The number of pixels is reduced from 1024 × 1024 (raw interferogram) to 152 × 152 (complex-field reconstruction) through the phase retrieval process. The reconstructed image has the diffraction-limited pixel size of ~500 nm. The visible illumination power at the sample plane can be increased up to ~100 μW.

**Phase canceling.** The phase canceling requires the following calibration for estimating the voltages to be loaded on each pixel of the SLM (consisting of N pixels) from the OPD image measured with DH (consisting of M pixels), where the number of pixels generally do not match between the two images with N > M. We make a set of calibration images that links the N-pixel SLM image and M-pixel OPD image for each of 256 (8 bit) phase gradients of the SLM. This can be made by inputting a uniform voltage to all the N pixels of the SLM and measuring the corresponding M-pixel OPD image by DH. Then, by using the set of calibration images we translate the measured M-pixel OPD image to the N-pixel SLM's input voltage image for phase canceling. In case the phase canceling does not sufficiently work, we can iteratively run the canceling procedure with feedback upon the uncanceled remaining OPD distribution.

**Samples.** The porous silica microbeads [43-00-503, Sicastar (micromod Partikeltechnologies GmBH)] immersed in index-matching oil (refractive index 1.50 at 587.56 nm, Shimadzu) is used as the sample for the experiments shown in Fig. 2 - 4. The COS7 cells (Riken) for the experiment shown in Fig. 5 are cultured in Dulbecco's Modified Eagle's Medium (DMEM) with 10% fetal bovine serum supplemented with penicillin–streptomycin, L-glutamine, sodium pyruvate and nonessential amino acids at 37 °C in 5% $CO_2$. For the live-cell imaging, the cells are cultured in a 35-mm glass-bottomed dish (AGC Techno Glass) and the medium is replaced by phenol red-free culture medium containing HEPES buffer (2 mL) before imaging. All solutions are from Thermo Fisher Scientific.

**MIR photothermal QPI of microbeads.** MIR pulses of a duration of 5 μs lasing at 1,045 $cm^{-1}$ provided by a quantum cascade laser (QD9500CM1, Throlabs) is used as the pump light. A ZnSe lens (LA7733-G, Thorlabs) with a focal length of 20 mm is used to loosely focus the MIR light to the sample with an excitation-field diameter of ~75 μm. The MIR ON-OFF modulation rate is 5 Hz. The MIR pulse energy at the sample plane is ~50 nJ. The diameter of the DF mask is 100 μm.

**MIR photothermal QPI of a COS7 cell.** MIR pulses of a duration of 1 μs lasing at 1,550 $cm^{-1}$ provided by a quantum cascade laser [DO418, Hedgehog (Daylight Solutions)] is used as the pump light. A ZnSe lens (LA7733-G, Thorlabs) with a focal length of 20 mm is used to loosely focus the MIR light to the sample with excitation-field elliptical major and minor axes of ~70 μm and ~30 μm, respectively. The MIR pulse energy at the sample plane is ~100 nJ. The diameter of the DF mask is 50 μm.

### Data availability
The data provided in the manuscript is available from T.I. upon request.

### Acknowledgements

We thank Haruyuki Sakurai for giving us feedback about the manuscript. We are grateful to Masaharu Takarada and Kohki Okabe for offering biological cells. This work was financially supported by JST PRESTO (JPMJPR17G2).

### Author contributions
K.T. conceived the concept, designed and constructed the systems, performed the experiments and analyzed the data. M.T. contributed in design of the systems and the experiments and data interpretation. T.I. supervised the entire work. All authors wrote the manuscript.

**Competing interests**

Authors declare no competing interest.

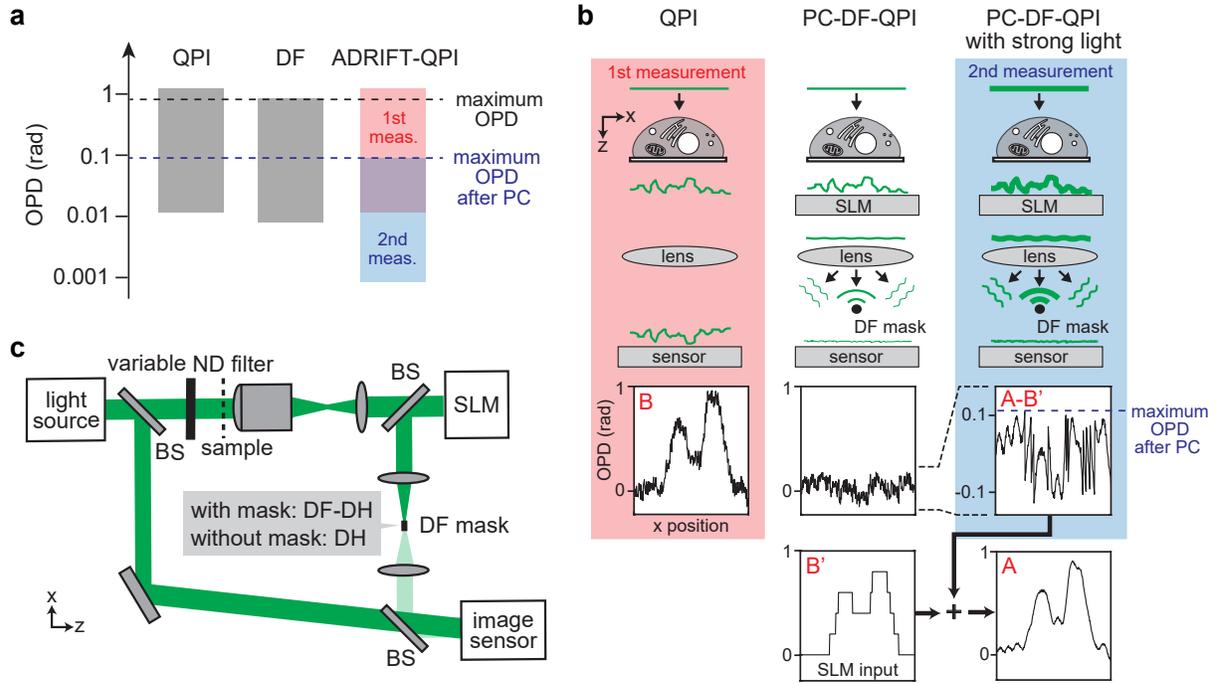

**Fig. 1 | Principle of ADRIFT-QPI. a,** Dynamic range of phase imaging: (left) conventional QPI, (center) dark-field imaging, (right) ADRIFT-QPI. We assume a standard CMOS image sensor with a full well capacity of ~10,000 e⁻/pixel is used. **b,** Principle of dynamic range expansion in ADRIFT-QPI. The left column shows the first measurement, where the sample's large OPD distribution (B) is measured by QPI with a plane wave illumination. The center column shows the situation of PC-DF-QPI where the DF mask blocks the unscattered light by phase canceling with the SLM. The right column shows the second measurement, which is PC-DF-QPI with strong light illumination, allowing for dynamic-range-shifted highly sensitive measurement. The dynamic-range-expanded OPD distribution of the sample (A) can be computationally reconstructed by combining the SLM's input OPD map (B') and the PC-DF-QPI measurement result (A-B'). **c**, Optical implementation of ADRIFT-QPI. In this work, off-axis DH is used as a QPI technique. DH and DF-DH are interchangeable in a single setup by insertion/removal of the DF mask. The phase-only SLM is put in the sample's conjugate plane for wavefront shaping, while the DF mask in the Fourier plane for spatially filtering the unscattered light. The illumination light intensity onto the sample can be changed with a neutral density (ND) filter placed in front of the sample. BS: beamsplitter.

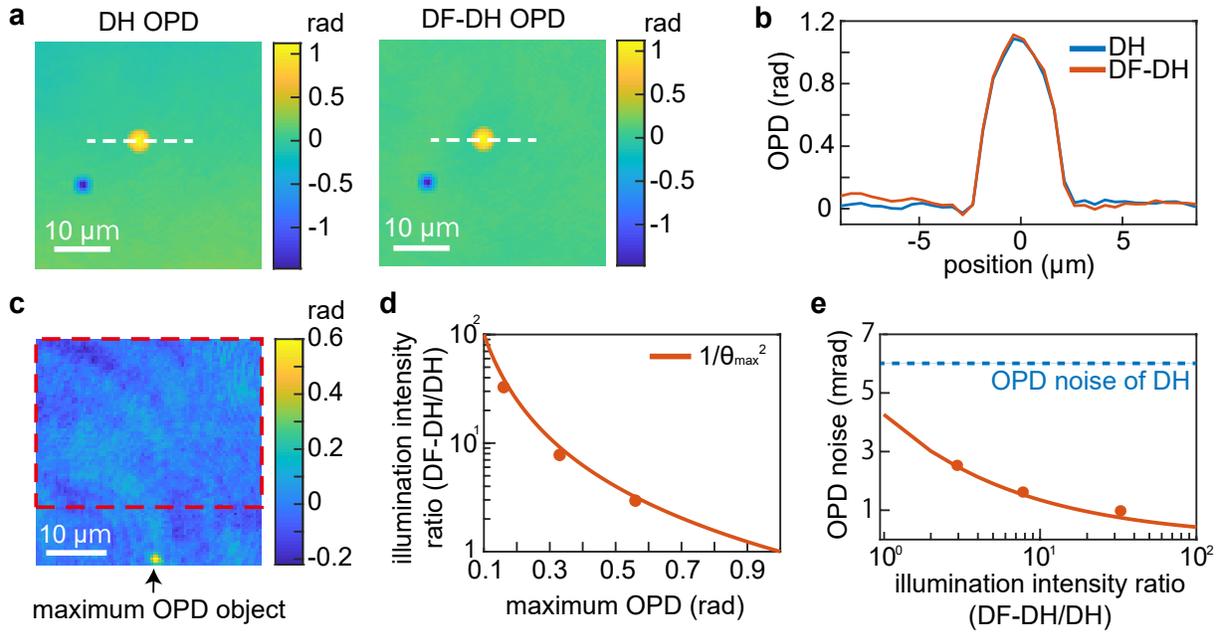

**Fig. 2 | Experimental validation of DF-DH. a,** OPD images measured by DH (left) and DF-DH (right). **b,** Cross-sectional profiles along the white dotted lines in **a**. The blue and orange curves represent the results obtained by DH and DF-DH, respectively. **c**, OPD image measured by DF-DH without the sample. An artificial object (indicated by the arrow) with the maximum OPD in the FOV (0.56 rad) is generated with the SLM. The area within the red rectangle is used for the sensitivity evaluation. **d,** Illumination intensity ratio of DF-DH and DH versus the maximum OPD value in the FOV. The data show good agreement with the theoretical curve (solid line). **e,** OPD noise (standard deviation of temporal OPD) versus illumination intensity ratio of DF-DH and DH. The blue dotted line indicates the OPD noise of DH and is shown as a reference. The data show good agreement with the theoretical curve (solid line).

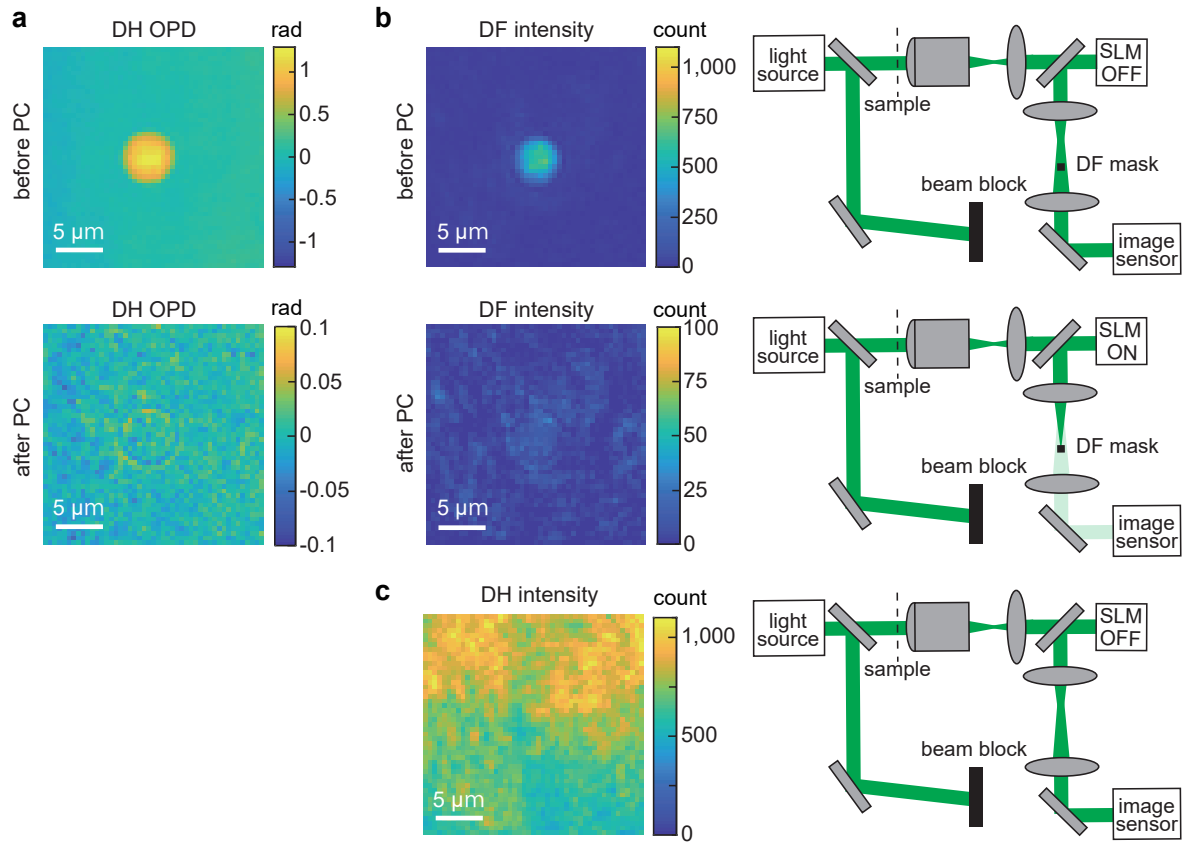

**Fig. 3 | Experimental validation of phase canceling. a,** DH OPD images before and after phase canceling. **b,** DF intensity images provided by the sample arm before and after phase canceling. **c,** DH intensity image provided by the sample arm.

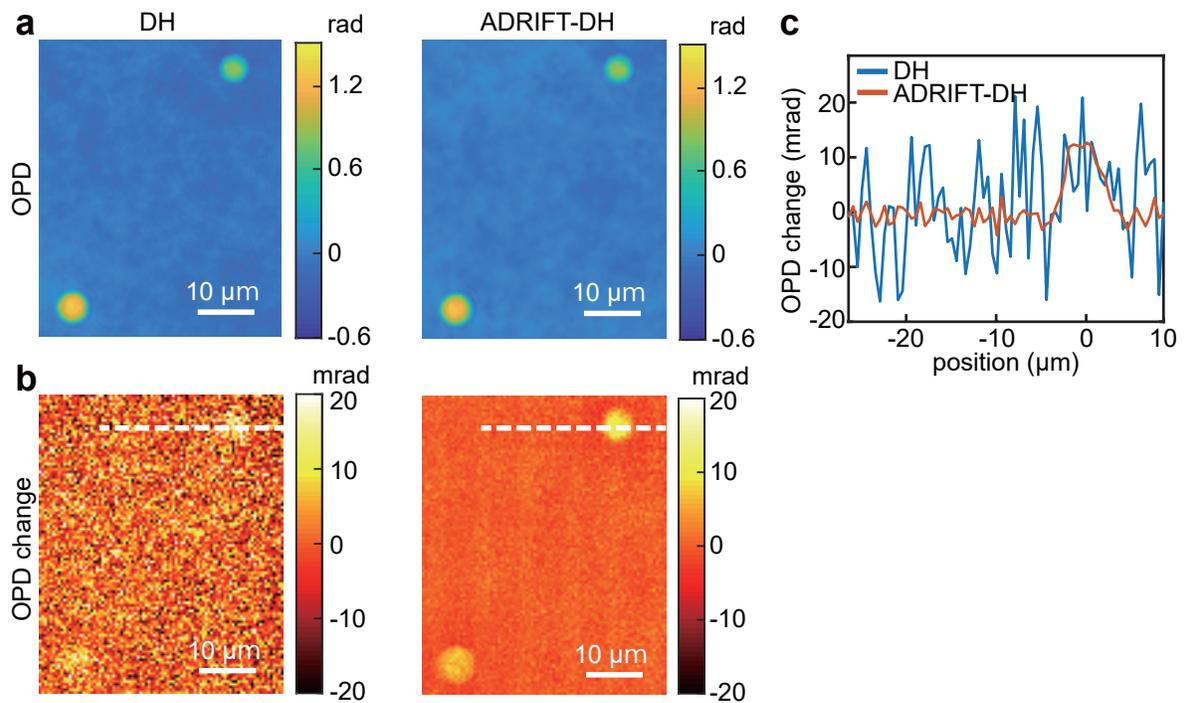

**Fig. 4 | Dynamic-range-expanded MIR photothermal QPI of silica microbeads. a,** OPD images measured by DH (left) and ADRIFT-DH (right) at MIR OFF state. **b,** Images of the photothermal OPD change due to absorption of the MIR pump light measured by DH (left) and ADRIFT-DH (right). **c,** Cross-sectional profiles along the white dotted lines in **b**. The blue and orange plots represent the results obtained by DH and ADRIFT-DH, respectively.

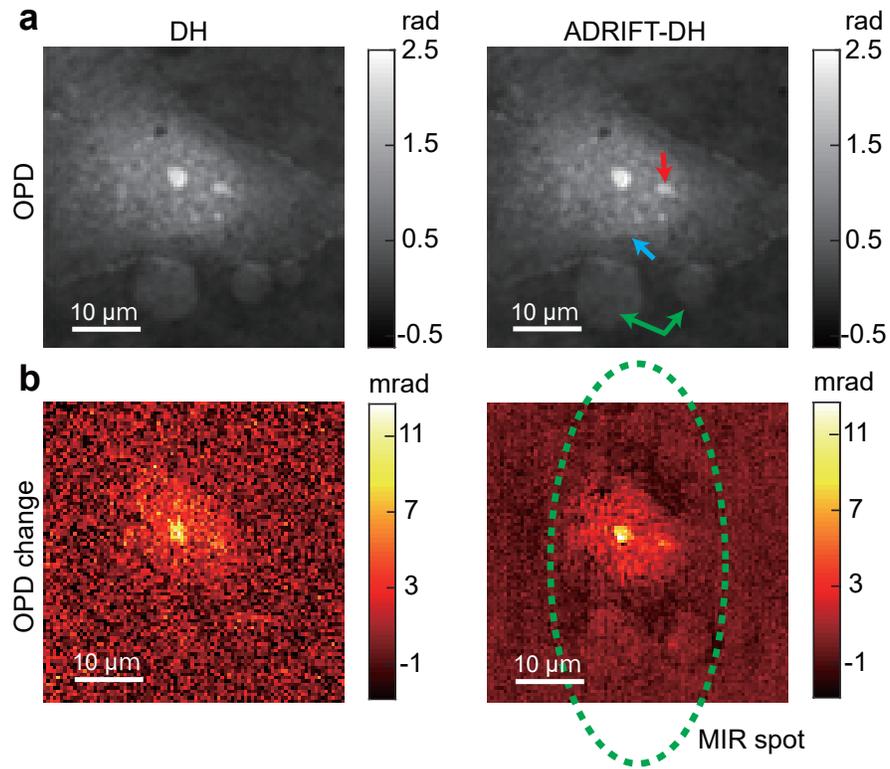

**Fig. 5 | Dynamic-range-expanded MIR photothermal QPI of a live COS7 cell. a,** OPD images measured by DH (left) and ADRIFT-DH (right) at MIR OFF state. The red, blue and green arrows show existence of the nucleoli, nucleus and two particles, respectively. **b,** Images of the photothermal OPD change due to absorption of the MIR pump light measured by DH (left) and ADRIFT-DH (right). The area in the green dotted circle indicates the illumination spot of the MIR pump light. The structures indicated by the red, blue and green arrows in **a** also give stronger signals in the photothermal images shown in **b**, which is visualized clearly with ADRFIT-DH but not with DH.